# Monte Carlo Study of Non-diffusive Relaxation of a Transient Thermal Grating in Thin Membranes

Lingping Zeng[1], Vazrik Chiloyan[1], Samuel Huberman[1], Alex A. Maznev[2],

Jean-Philippe M. Peraud[1], Nicolas G. Hadjiconstantinou[1], Keith A. Nelson[2],

and Gang Chen[1,a]

[1]Department of Mechanical Engineering, Massachusetts Institute of Technology,

Cambridge, Massachusetts 02139, United States

[2]Department of Chemistry, Massachusetts Institute of Technology,

Cambridge, Massachusetts 02139, United States

ABSTRACT:

The impact of boundary scattering on non-diffusive thermal relaxation of a transient grating in thin membranes is rigorously analyzed using the multidimensional phonon Boltzmann equation. The gray Boltzmann simulation results indicate that approximating models derived from previously reported one-dimensional relaxation model and Fuchs-Sondheimer model fail to describe the thermal relaxation of membranes with thickness comparable with phonon mean free path. Effective thermal conductivities from spectral Boltzmann simulations free of any fitting parameters are shown to agree reasonably well with experimental results. These findings are important for improving our fundamental understanding of non-diffusive thermal transport in membranes and other nanostructures.

[a] To whom correspondence should be addressed. Electronic mail: gchen2@mit.edu

Laser based thermal transport measurements at characteristic lengths comparable with phonon mean free paths (MFPs) have emerged as useful tools for determining phonon MFP distributions[1–9] in materials. Among them, transient thermal grating (TTG) spectroscopy probes the non-diffusive relaxation of a sinusoidal thermal grating imposed on a thin membrane[10,11]. In this process, as schematically depicted in Fig. 1, the sinusoidal thermal grating is formed through interference of two pump laser beams crossed at an angle. The diffraction decay from the thermal grating is measured by another probe beam and is subsequently fit to the heat diffusion theory to extract the effective thermal conductivity. The transport regime in the membrane is determined by the grating period λ, membrane thickness d, and the bulk MFPs of the membrane material. In the diffusive regime where the grating period is much longer than the MFPs of thermal phonons, the thermal decay follows an exponential profile $T(x,t) \propto \cos(qx)\exp(-\gamma t)$ according to the Fourier heat diffusion theory, with the decay rate $\gamma$ given by $\gamma = \alpha q^2 = \frac{kq^2}{C}$, where $q = \frac{2\pi}{\lambda}$ is the grating wavevector, $\alpha$ is the thermal diffusivity defined as the ratio of the thermal conductivity $k$ to the volumetric heat capacity $C$[4]. In the non-diffusive regime where the grating period is comparable with or shorter than the bulk MFPs of thermal phonons, long-MFP phonons do not scatter when they traverse a half period of the thermal grating. Consequently, the thermal transport deviates from the prediction of heat diffusion theory and the thermal decay rate does not follow the expected quadratic dependence on the grating period[4]. An effective thermal conductivity is defined in the non-diffusive regime based on the effective thermal decay rate, i.e. $k_{\text{eff}} = \frac{\gamma_{\text{eff}} C}{q^2}$. Here, we would like to emphasize the subtle distinction between the diffusive vs. non-diffusive transport compared to past extensive studies on size effects, especially classical size effects in thin films[12–14]. It has been well-established that when the film thickness is comparable or smaller than the phonon MFP, the measured effective thermal conductivity is suppressed due to phonon boundary scattering at the surfaces or interfaces. For heat conduction along the film plane direction, however, the Fourier law can still be applied as long as the film length is longer than the phonon MFP, despite that boundary scattering leads to a smaller effective thermal conductivity, and hence transport can still be considered diffusive at a coarse-grained level. In a TTG experiment, this diffusive transport is reflected in thermal diffusivity measurements that are independent of the grating period for a specific thin film. In the non-diffusive regime, the measured thermal diffusivity becomes dependent on the grating period. Non-diffusive transport behavior at micrometer lengthscales was recently observed in TTG measurements using a 390 nm thick crystalline silicon membrane[4].

Although phonon boundary scattering plays an important role in the non-diffusive relaxation of transient gratings in membrane samples, prior theoretical studies invoking the phonon Boltzmann transport equation (BTE) typically assumed transport in one spatial dimension and neglected the impact of the phonon boundary scattering in the membrane[15–18]. Minnich extended the 1D model to a 2D semi-infinite body system using an analytical Green's function method[19]. However, the Green's function approach is not applicable for modeling thermal transport in thin membranes and is also restricted to specular boundary scattering that is not appropriate for thin membranes.

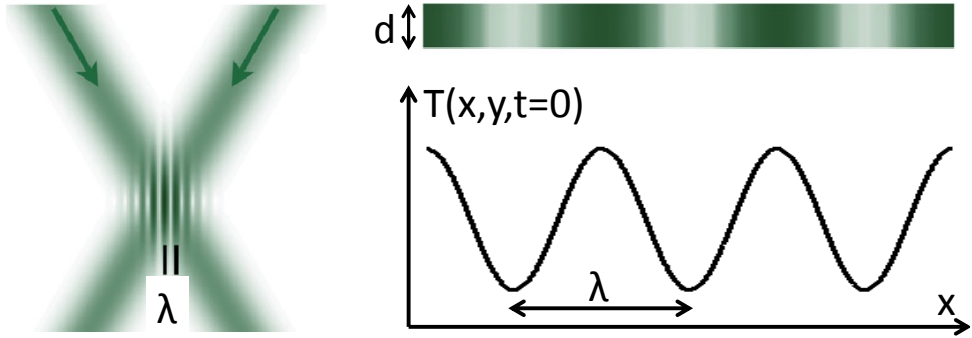

Figure 1. Schematic of the membrane geometry used in the TTG experiment (adapted from Ref. 13). The interference pattern from crossing two pump laser beams generates a spatially sinusoidal temperature profile at $t = 0$ along the membrane. The decay of the thermal grating profile depends on the grating period $\lambda$, the membrane thickness d, and the phonon MFPs in the membrane.

In this work, we study the non-diffusive thermal relaxation in thin membranes used in TTG experiments by rigorously solving the 2D phonon BTE with diffuse scattering at the membrane boundaries. We assume completely diffuse scattering as the use of diffuse boundary scattering in most studies gives satisfactory results[20–22]. Note that diffuse boundary scattering does not preclude thermal transport from becoming non-diffusive at ultrashort transport distances, as shown in prior work[4] and this work. Modeling frameworks based on both gray and spectral phonon BTE are developed with the goal of examining the impact of phonon boundary scattering on the non-diffusive thermal transport. The gray model is used for finding heat flux suppression functions (a measure of reduction in phonon heat flux from different-MFP modes for a given material system relative to the diffusion prediction) appropriate for the experimental geometries to approximately extract the MFP distribution[15,16,23]. The first-principles based spectral model is capable of simulating the experimental observable and modeling non-diffusive thermal relaxation in the TTG experiments without any fitting parameters. It rigorously accounts for the phonon spectra from first-principles density functional theory (DFT) calculation to obtain the size-dependent thermal conductivities that are subsequently compared with reported experimental results.

Due to the computational cost associated with using deterministic approaches for solving the BTE[24], here we use a recently developed deviational Monte Carlo (MC) technique by Peraud and Hadjiconstantinou[25,26] that significantly reduces the stochastic noise and improves the computational efficiency by simulating only the deviation from a reference equilibrium distribution. Given the small temperature rise associated with TTG experiments ($O(1K)$), we use the kinetic version of the algorithm[22] that is significantly more efficient and applicable when the BTE solution can be linearized about a reference equilibrium. The details of the technique are described elsewhere[26–28] and are only briefly discussed here. The simulation technique owes much of its computational efficiency to the fact that it can treat computational particles independently and thus sequentially. The number of computational particles to be simulated is determined by balancing the computational cost with statistical uncertainty. The simulation of each particle, representing a fixed amount of energy, starts by initializing the particle's initial frequency and polarization from the appropriate frequency distribution[22]; the initial spatial location of the particle is drawn from an appropriately normalized distribution reflecting the initial temperature distribution, in our case a spatially sinusoidal profile along the in-plane direction (uniform across the membrane thickness). The simulation proceeds by advancing each particle's trajectory forward in time as a sequence of straight-line advection events interrupted by scattering events. The time between two anharmonic scattering events is calculated by drawing a random number from the exponential distribution with a decay rate $\tau^{-1}$, where $\tau$ is the relaxation time of the particle under consideration. If during advection a particle encounters either boundary of the membrane, the traveling direction of the particle is randomized, as required by the diffuse scattering boundary condition. If no boundary is encountered up to the time of the next anharmonic scattering event, the simulation proceeds by processing the scattering event by resetting the particle's frequency, polarization and traveling direction (only the traveling direction for the gray model)[29]. We note that although the thermal transport is two-dimensional, phonons travel in three dimensions. The simulations take into account this by generating three travelling directions ($k_x$, $k_y$, $k_z$) for each individual computational particle[30]. When either boundary or internal scattering occurs, all three travelling directions are appropriately reset, ensuring correct outgoing probabilities over the solid angle space. The simulation ends when the desired simulation time is reached. The temporal decay of the peak-valley temperature difference, corresponding to the measured diffraction signal in the TTG experiments, is sampled from the recorded trajectories of the computational particles. The effective thermal conductivity is obtained by matching the heat diffusion theory solution to the thermal decay from the MC simulation[4]. We would like to clarify that our MC approach[26] is different from the phonon MFP sampling method used by McGaughey and Jain[31] which is an approximate method for sampling MFPs and cannot be used to simulate dynamic behavior.

In the gray model of the thermal relaxation process, two nondimensional parameters are important: the Knudsen number $\text{Kn} = \frac{\Lambda}{d}$ and the ratio of the MFP to the grating period $\eta = \frac{2\pi\Lambda}{\lambda}$, where $\Lambda$ is the gray phonon MFP. The Knudsen number describes the lateral size effect caused by the finite membrane thickness along the cross-plane direction[13] and $\eta$ describes the non-diffusive effect caused by the finite transport length along the in-plane membrane direction[15–17]. We simulate the thermal relaxation for the geometry of Fig. 1 across a wide range of Kn and $\eta$ by solving the gray MC model using a single set of phonon properties (including heat capacity, group velocity, and MFP) for a wide range of grating periods and membrane thicknesses. Since $\eta$ is the only variable that affects the thermal relaxation for a given Kn, varying the grating period has the same effect as varying the phonon MFP.

Representative BTE solutions in the thick membrane and large grating period limits are shown in Figs. 2(a) and 2(b), respectively. Diffusion model fits to these solutions used to find the heat flux suppression function $S_{\text{gray}}(\eta, \text{Kn}) = k_{\text{eff}}/k_{\text{bulk}}$, where $k_{\text{eff}}$ and $k_{\text{bulk}}$ are the effective and bulk thermal conductivities, respectively, are also included. The suppression function describes the thermal conductivity reduction caused by non-diffusive transport and is discussed in detail later[16,23]. In the thick film limit (Fig. 2(a), obtained for Kn = 0.01), thermal transport transitions from diffusive to ballistic when $\eta$ is varied from 0.1 to 10, consistent with prior 1D BTE modeling results[16]. In the diffusive regime ($\eta \sim 0.1$), the thermal decay is exponential and the thermal conductivity is well defined based on the decay rate. The thermal relaxation in the ballistic regime ($\eta > 1$) becomes non-exponential and cannot be accurately described by the heat diffusion theory, as expected. However, it is still useful to fit the BTE solution with the diffusion theory to obtain an effective thermal conductivity which gives a measure of the thermal relaxation rate relative to the diffusion prediction. In fact, this strategy is commonly used in the analysis of non-diffusive thermal transport in the heat transfer field[1,3,4,6,9,32]. One important benefit of this strategy is that the intrinsic phonon MFP distribution can be approximately extracted from the length-dependent effective thermal conductivities since Fourier's law is used consistently in the data analysis[6,7,9,16,23]. We emphasize that the effective thermal conductivity is geometry specific, meaning that the effective thermal conductivity obtained in this study is only valid for the transient grating geometry and cannot be used for other experimental configurations. In the ballistic regime, the effective thermal conductivity is generally much lower than the bulk value, indicating a much slower decay dynamic compared to the diffusion prediction.

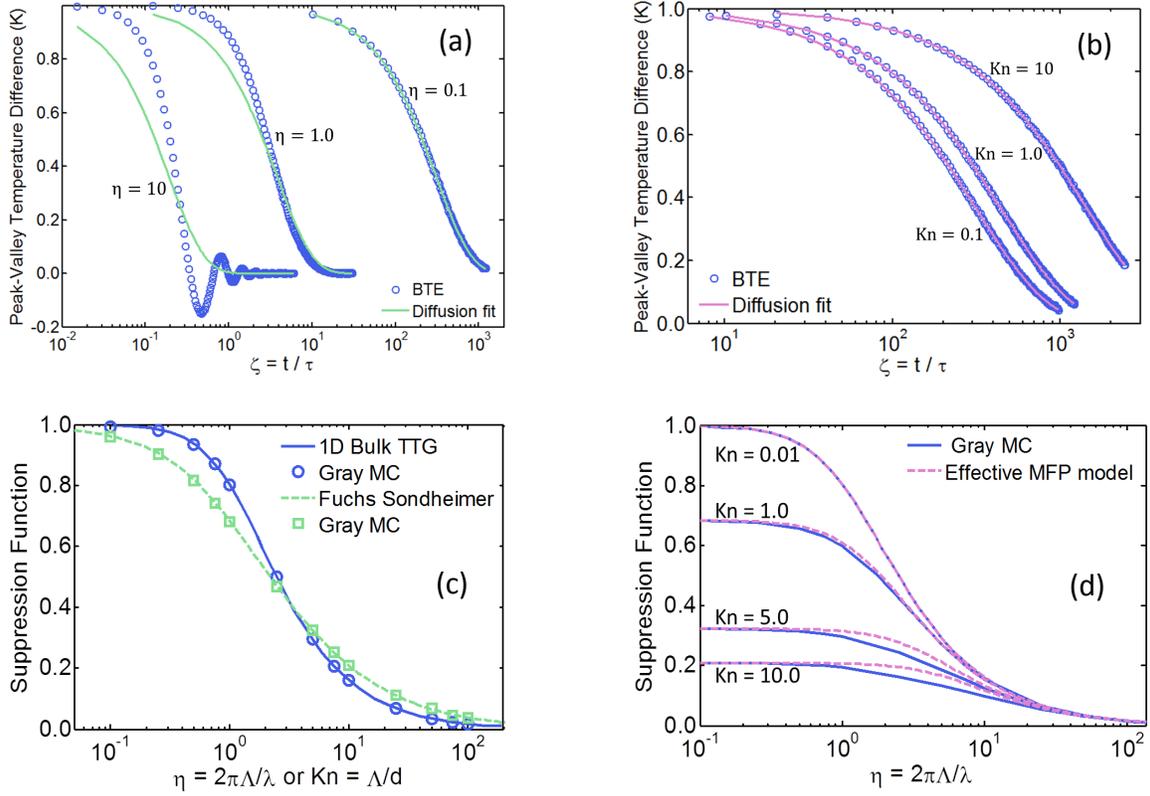

Figure 2. (a) Representative peak-valley temperature decay traces in the thick membrane limit: Kn = 0.01. (b) Representative peak-valley temperature decay traces in the large grating period limit: η = 0.1. (c) Comparison between MC calculated suppression functions with the 1D bulk TTG limit for thick membranes (Kn = 0.01) and Fuchs Sondheimer model for large grating periods (η = 0.1). (d) Computed suppression

functions versus η across a range of Kn. The suppression function based on the effective MFP model are given by $S_E(\eta, \text{Kn}) = S_{\text{1DTTG}}(\eta S_{\text{FS}}(\text{Kn})) \times S_{\text{FS}}(\text{Kn})$.

In the large grating spacing limit (Fig. 2(b), obtained for η = 0.1), lateral size effects occur when the membrane thickness becomes comparable with the phonon MFP. The normalized effective thermal conductivities decrease substantially with increasing Kn due to increasing phonon boundary scattering as the membrane thickness is reduced. However, transport in this limit can still be viewed as diffusive regardless of the membrane thickness, as shown by the excellent fits based on diffusion theory. This is consistent with the Fuchs-Sondheimer model[33,34] of thin film thermal conductivity which neglects the presence of the thermal grating, since lateral size limitation is perpendicular to the heat flux direction. Figure 2(c) compares the heat flux suppression functions from the MC simulations in these two limits with prior numerical results from Collins *et al.*[16], denoted by $S_{\text{1DTTG}}(\eta)$, and the Fuchs-Sondheimer model, denoted by $S_{\text{FS}}(\text{Kn})$[33,34], respectively. The excellent agreement validates the accuracy of the MC model.

For membrane thickness and grating period comparable with the phonon MFP, the lateral size effect and the non-diffusive transport along the heat flux direction come into play simultaneously. Figure 2(d) compares the MC calculated suppression functions $S_{\text{gray}}(\eta, \text{Kn})$ as a function of η across a range of Kn with one approximating model. The approximating model, referred to as the 'effective MFP model' (originally used in Ref. 4), treats the effective phonon MFP in the membrane as the bulk MFP multiplied by the suppression factor given by the Fuchs-Sondheimer model, i.e. $\Lambda_{\text{eff}} = \Lambda \times S_{\text{FS}}\left(\frac{\Lambda}{d}\right)$. The suppression function derived from the effective MFP model is given by:

$$S_E(\eta, \text{Kn}) = S_{\text{1DTTG}}\left(\frac{2\pi\Lambda_{\text{eff}}}{\lambda}\right) \times S_{\text{FS}}\left(\frac{\Lambda}{d}\right) = S_{\text{1DTTG}}(\eta S_{\text{FS}}(\text{Kn})) \times S_{\text{FS}}(\text{Kn}) \qquad (1)$$

As shown in Fig. 2(d), increasing Kn results in decreasing thermal conductivity of the membrane due to stronger lateral size effects caused by the membrane boundaries. In the thick film limit (small Kn), the effective MFP model agrees well with the MC simulation result. When the membrane thickness becomes comparable or smaller than the phonon MFP (Kn > 1.0), the effective MFP model underestimates the suppression effect due to its underestimate of the MFP in the membrane. This highlights the importance of rigorously accounting for the impact of boundary scattering on the thermal transport in real thin membranes.

Now let us consider the spectral MC model to describe the non-diffusive thermal relaxation of transient grating in real membranes. This model rigorously takes into account the phonon dispersion and frequency dependent lifetimes from first-principles DFT calculations[35] to simulate the experimental observable (i.e. the diffraction decay) for real membrane samples without any adjustable parameters. The goal is to examine whether the first-principles based simulations can match the experimental results. To demonstrate its viability, we apply the spectral model to study the thermal relaxation of crystalline silicon membranes across a wide range of grating periods and thicknesses. Representative decay profiles from the spectral MC simulation and the corresponding diffusion model fits are shown in Fig. 3(a) for the experimentally studied 390 nm thick silicon membrane[4]. For the experimentally achievable grating periods ($\lambda > 1$ μm), the thermal transport is in the weakly non-diffusive regime, as can be seen from the excellent fits in Fig. 3(a) based on the diffusion theory with a modified thermal conductivity. It is only for $\lambda \ll 1$ μm, whereby the grating period becomes much smaller than the MFPs of dominant thermal phonons, that transport becomes ballistic (see Fig. 3(a) for 10 nm and 100 nm grating periods). In fact, even in the ballistic regime, the oscillatory feature of the strongly ballistic behavior is not observed. We emphasize that it is the finite heat transport length imposed by the transient thermal grating along the membrane that leads to the observed non-diffusive transport. The phonon boundary scattering due to the finite membrane thickness is not the cause of non-diffusive transport defined here, as demonstrated previously in Fig. 1(b) and also in prior experimental measurement on silicon membranes across a range of thicknesses and grating periods[4,7]. A thermal conductivity, smaller than the bulk value, can still be defined for the membrane (as given, for example, by the Fuchs-Sondheimer model) in the presence of only phonon boundary scattering (including partially specular boundary scattering). However, the ballistic transport induced by the finite grating period significantly lowers the measured thermal conductivity values below the effective value of the film itself.

Normalized effective thermal conductivities as a function of grating period for silicon membranes across a range of thicknesses are shown in Fig. 3(b). The effective thermal conductivity decreases with decreasing grating period, indicating a stronger non-diffusive effect for shorter grating period. As shown in Fig. 3(b), the onset of non-diffusive thermal transport shifts to shorter grating period when the membrane thickness becomes smaller, consistent with shorter effective MFPs due to stronger boundary scattering for thinner membranes. In the limit of extremely short grating period, the non-diffusive effect due to the finite transport length dominates and the thermal conductivity becomes independent of the membrane thickness. In contrast, in the limit of large grating period, thermal transport approaches the diffusive regime and the thermal conductivity consequently approaches the effective value given by the Fuchs-Sondheimer model.

Figure 3(c) compares the normalized effective thermal conductivities from the spectral model with the experimental data from Ref. 4. Also included in Fig. 3(c) are the 'forward' calculated effective thermal conductivities based on the gray MC suppression function $S_{gray}(\eta, Kn)$ and the effective MFP model $S_E(\eta, Kn)$. These calculations are referred to as 'forward' to highlight their difference with 'inverse' problems where effective thermal conductivities and relevant suppression functions are known to reconstruct the MFP distributions. In these 'forward' calculations, the effective thermal conductivity is related to the bulk MFP distribution through the suppression function by the relation[15,16,23]:

$$k_{eff}(\lambda, d) = \int_0^\infty S\left(\frac{2\pi\Lambda}{\lambda}, \frac{\Lambda}{d}\right) f(\Lambda) d\Lambda = \int_0^\infty K\left(\frac{2\pi\Lambda}{\lambda}, \frac{\Lambda}{d}\right) F(\Lambda) d\Lambda \qquad (2)$$

where $f(\Lambda)$ is the differential thermal conductivity per MFP, $F(\Lambda)$ is the cumulative thermal conductivity distribution function, and $K\left(\frac{2\pi\Lambda}{\lambda}, \frac{\Lambda}{d}\right) = -\frac{dS}{d\Lambda}$ is the kernel function.

$F(\Lambda)$ describes the fractional contribution of phonons with MFPs shorter than a threshold value $\Lambda$ to thermal conductivity[36,37]: $F(\Lambda) = \int_0^\Lambda f(\Lambda')d\Lambda'$. The key assumption in Eq. (2) is that the suppression function depends only on the sample geometry and the dependence of the effective thermal conductivity on the material properties occurs only through the thermal conductivity accumulation function[9,23]. The benefits of this assumption are two-fold. On one hand, this assumption allows the extraction of intrinsic phonon MFP distributions from size-dependent thermal conductivities without any prior knowledge of phonon scattering processes in the material, as shown by prior studies[6,7,9,16,23]. On the other hand, it allows the suppression function to be computed from the gray-body BTE since the same suppression function applies to a model gray medium in which all the phonons have the same MFP. Although this assumption is not completely rigorous, it has been shown to work reasonably well in describing thermal transport along one-dimensional transient thermal grating[16] and thermal transport across thin films[38]. While realizing its limitations, we use this assumption in the computation of the 'forward' thermal conductivities with the DFT phonon spectra input to examine how well this assumption describes the thermal relaxation of transient grating in membranes.

As shown in Eq. (2), for a given membrane thickness d and grating period λ, the gray MC suppression functions depend only on the phonon MFP. Figure 3(d) shows the resulting gray suppression functions for a 390 nm thick membrane across a range of grating periods. Stronger suppression effect occurs at shorter grating period due to stronger non-diffusive effect caused by shorter transport length along the membrane. When the grating period becomes large, the suppression function approaches the Fuchs Sondheimer limit, as expected. The 'forward' calculations use the differential thermal conductivity from first-principles DFT calculations[35] and the gray suppression functions to compute the grating-period dependent effective thermal conductivity. As shown in Fig. 3(c), the experiment measured a relatively flat effective thermal conductivity above 7.5 μm grating period, indicating near-diffusive thermal transport for large heat transfer lengths. However, a sharp reduction of the effective thermal conductivity was observed for grating periods below 5 μm, implying strong non-diffusive transport effect. None of the model predictions agree exactly with the measurement results. The disparity between the spectral model and the experiment above 7.5 μm grating period may arise from the finite boundary specularity of the membrane samples. In the strongly non-diffusive regime, the spectral model prediction agrees slightly better with the experimental result compared with the 'forward' calculations using the gray MC suppression function and the effective MFP model, indicating the importance of accounting for the phonon dispersion and spectral lifetimes in the simulation.

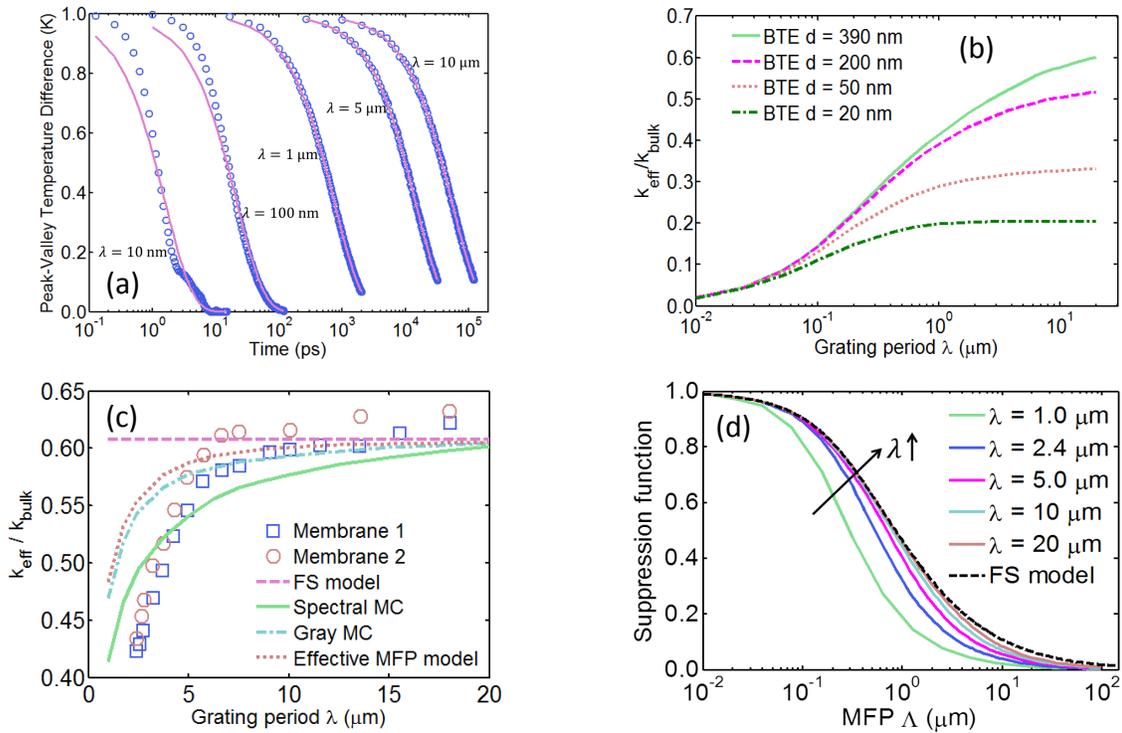

Figure 3. (a) Representative peak-valley temperature decay traces from spectral MC simulation (dots) and diffusion fits (solid lines). (b) Normalized effective thermal conductivities for silicon membranes of four different thicknesses from the spectral MC calculations. (c) Normalized effective thermal conductivities from experiments (Ref. 4), spectral MC simulation, forward calculations using the gray MC suppression functions

and the effective MFP model. The Fuchs Sondheimer limit (dashed line) is plotted as a reference. (d) Computed gray suppression functions versus phonon MFP across a range of grating period for a 390 nm thick membrane.

In summary, we study the impact of phonon boundary scattering on the non-diffusive relaxation of a transient thermal grating in thin membranes used in TTG spectroscopy experiments by rigorously solving the phonon Boltzmann transport equation. Our gray BTE model results show that approximating models cannot accurately describe the thermal relaxation when the membrane thickness becomes comparable with phonon MFPs. The length-dependent thermal conductivities from the spectral BTE model agree reasonably well with the measured data on a 390 nm silicon membrane for grating periods longer than 1 μm. These results help gain insight into fundamental understanding of non-diffusive thermal transport physics in transient grating spectroscopy and are also important for understanding non-diffusive thermal behavior in general nanostructures.

The authors are grateful for helpful discussions with Dr. Kimberlee C. Collins. This material is based upon work supported as part of the "Solid State Solar-Thermal Energy Conversion Center (S3TEC)", an Energy Frontier Research Center funded by the U.S. Department of Energy, Office of Science, Office of Basic Energy Sciences under Award Number: DE-SC0001299/DE-FG02-09ER46577.